\documentclass[]{elsart3p}

\usepackage{graphicx}

\usepackage{amssymb}

\usepackage{lineno}

\newcommand{\rmi}{\mathrm{i}}

\begin{document}

\begin{frontmatter}



\title{Microwave vortex dynamics  in Tl-2212 thin films}


\author[RomaTre]{N. Pompeo\corauthref{cor}},
\corauth[cor]{Corresponding author.}
\ead{\\pompeo@fis.uniroma3.it}
\author[IPHT]{H. Schneidewind},
\author[RomaTre]{E. Silva}
\address[RomaTre]{Dipartimento di Fisica ``E. Amaldi'' and Unit\`a CNISM,
Universit\`a Roma Tre, Via della Vasca Navale 84, 00146 Roma,
Italy}

\address[IPHT]{Institute of Photonic Technology, P.O.B.100239, D-07702 Jena, Germany}

\begin{abstract}
We present measurements of the effective surface impedance changes due to a static magnetic field, $\Delta Z(H,T)=\Delta R(H,T)+\rmi \Delta X(H,T)$, in a Tl-2212 thin film with $T_c>$ 103 K, grown on a CeO$_2$ buffered sapphire substrate. Measurements were performed through a dielectric resonator operating at 47.7 GHz, for temperatures 60 K$\leq T<T_c$ and magnetic fields $\leq0.8$ T. We observe exceptionally large field induced variations and pronounced super-linear field dependencies in both $\Delta R(H)$ and $\Delta X(H)$ with $\Delta X(H)>\Delta R(H)$ in almost the whole $(H,T)$ range explored. A careful analysis of the data allows for an interpretation of these results as dominated by vortex dynamics. In the intermediate-high field range we extract the main vortex parameters by resorting to standard high frequency model and by taking into proper account the creep contribution.
The pinning constant shows a marked decrease with the field which can be interpreted in terms of flux lines softening associated to an incipient layer decoupling. Small vortex viscosity, by an order of magnitude lower than in Y-123 are found. Some speculations about these findings are provided.

\end{abstract}

\end{frontmatter}

\section{Introduction}
\label{intro}

The softness of the fluxon system in high-$T_c$ superconductors (HTS), favoured by anisotropy, small coherence length and large penetration length, makes it strongly susceptible to disorder induced by both thermal effects and order parameter inhomogeneities. An extreme limit of fluxon line softening shows up in layered compounds, where vortex lines leave place to 2D segments (pancakes) moving independently.
Consequently, the (H,T) plane is populated by very heterogenous vortex matter phases differing under many aspects, among which pinning mechanisms are of particular interest.

Vortex motion induced by a transport current is the main source of power dissipation. It is influenced by the nature of pinning and of the vortex system, which together determine the vortex mobility \cite{blatterone}. Moreover, the vortex motion dissipation involves quasi-particles (QP) excitations, whose density of states can be both bound inside or extending outside the vortex cores (as in nodal superconductors). In this sense the vortex motion can indirectly probe the electronic state inside vortices \cite{tsuchiya}.

In these respects the microwave response is a powerful probe: the small amplitude of the vortex oscillations induced at microwave frequencies reduces the effects of the complex interaction between vortices in the dynamic behaviour. The simple, single-vortex elastic regime is often excited by microwaves, which is represented by the so-called pinning constant $k_p$. The power dissipation is represented through the vortex viscosity $\eta$, which is linked to the vortex core properties in terms of QP density of states and lifetime. Additional extrinsic effects that might arise from grain boundaries and, in general, weak-links, are often identified by their peculiar magnetic field dependence.
Within standard Abrikosov vortex dynamics, the microwave response has been modeled by many authors \cite{GR,CC,brandt}. According to Brandt's approach \cite{brandt}, in the limit of negligible QP conductivity with respect to superfluid conductivity ($\sigma_1\ll\sigma_2$, i.e. $T\ll T_c$) the following expression can be written down:
\begin{equation}
\label{eq:rhoc}
    \tilde{\rho}=\rho_{vm}+\rmi\frac{1}{\sigma_2}
    =\rho_{ff}\frac{\epsilon'+\rmi\omega\bar{\tau}}{1+\rmi\omega\bar{\tau}}+\rmi\frac{1}{\sigma_2}
\end{equation}
\noindent where the second equality defines the vortex motion resistivity $\rho_{vm}$ and
%
$\rho_{ff}=\Phi_{0}B/\eta$ is the flux flow resistivity, $\bar{\tau}=\frac{\tau\tau_{r}}{\tau+\tau_{r}}$, where $\tau=\eta/k_p$ is the depinning characteristic time and $\tau_{r}=\tau e^{U/K_BT}$ is the creep relaxation time in the linear regime \cite{brandt}. Finally, $U$ is the pinning potential barrier height and $\epsilon'=\frac{\tau}{\tau+\tau_{r}}$ is a dimensionless creep parameter. In the limit of no creep $U\rightarrow\infty$, $\epsilon'\rightarrow0$, $\bar{\tau}\rightarrow\tau$ and $\rho_{vm}$ in Equation (\ref{eq:rhoc}) reverts to the simpler Gittleman-Rosenblum (GR) expression \cite{GR}. By increasing the excitation frequency, one spans from a pinning to a dissipation dominated dynamics which, for sufficiently high frequencies, becomes pure flux flow. The crossover is given by the characteristic (de)pinning angular frequency $\omega_p=k_p/\eta=1/\tau$. It is however worth noting that the above described model does not apply in full to the case of large creep rates (e.g., when $T$ approaches $T_c$), so that an estimate of the creep factor $\epsilon'$ is necessary in the analysis of the data.

In the following, we focus on the Tl-2212 compound, which has received little attention with respect to other cuprates such as YBCO and BSCCO. Moreover, it has an intermediate anisotropy between the more isotropic Y-123 and the strongly layered Bi-2212, so that its vortex phase properties can be a bridge between the almost rigid, rod like fluxons in Y-123 and the pancake behaviour of Bi-2212 fluxons.

\section{Experimental results and discussion}
\label{exp}
\begin{figure}[h]
\centerline{\includegraphics[width=8.5cm]{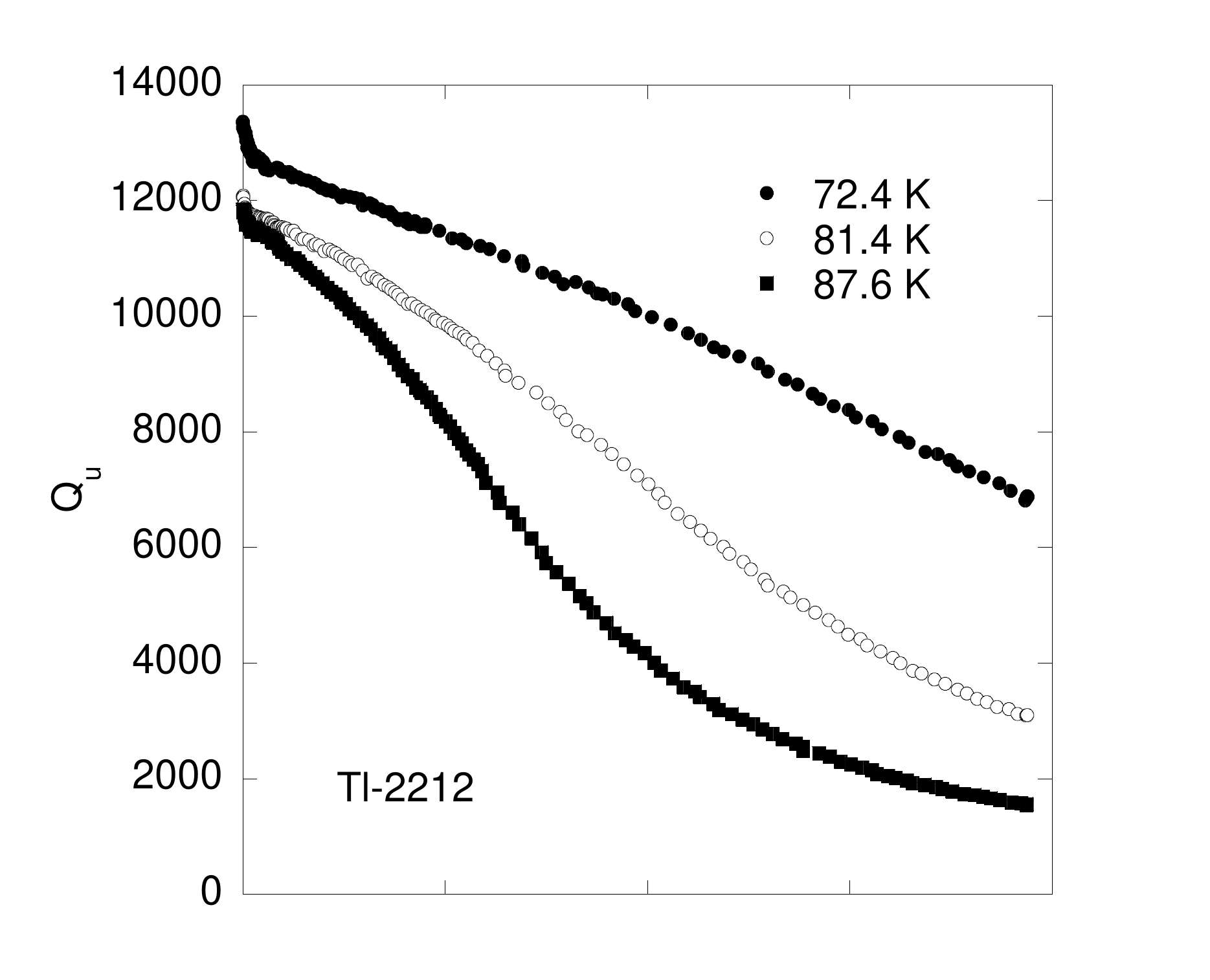}}
\vspace{-5mm}
\centerline{\includegraphics[width=8.5cm]{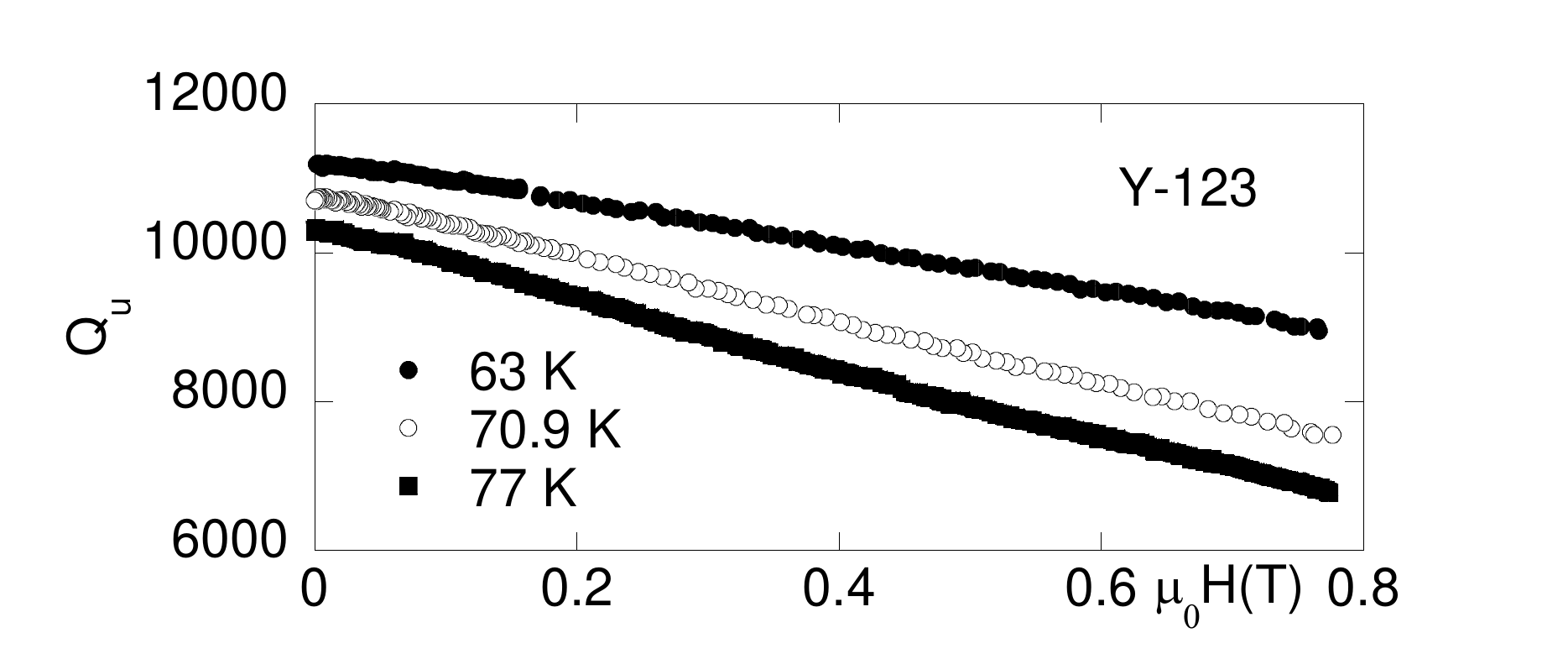}}
\vspace{-3mm}
  \caption{{\it Upper panel}: $Q_u vs. H$ at selected temperatures for Tl-2212. {\it Lower panel}: analogous measurements on Y-123, reported for same values of reduced temperatures $T/T_c$.}
\label{fig:Qu}
\end{figure}
Measurements are performed on a Tl-2212 thin film square sample, 10 mm wide and $d$=240 nm thick. It was grown c-axis-oriented (i.e. with the film c-axis parallel to the substrate normal) on CeO$_{2}$-buffered sapphire 0.44 mm thick by conventional two step method \cite{schneidewind}. The critical temperature, estimated from the microwave response, was $T_c>$ 103 K. The inductively measured critical current density is $J_c$(77 K)=0.5 MA/cm$^2$. A solid/liquid nitrogen cryostat provides temperatures down to 60 K. A conventional electromagnet generates magnetic fields $\mu_0H\leq0.8$ T, applied perpendicular to the sample surface (i.e. parallel to the sample $c$-axis). The microwave response is determined by means of a cylindrical sapphire resonator using the TE$_{011}$ mode with the surface perturbation method, at a resonating frequency $\omega/(2\pi)\approx47.7$ GHz. The resonator unloaded quality factor $Q_u$ and $\omega$ are determined by a Lorentz fit of the measured microwave power reflected by the resonator as a function of frequency \cite{dielRes}. Their field induced changes yield the field changes of the effective surface impedance $\Delta Z(H)=Z(H)-Z(0)=\Delta R(H)+\rmi \Delta X(H)$ according to:
\begin{equation}
\label{eq:measrho}
    \Delta Z(H)=G_s\left(\Delta\frac{1}{Q_u(H)}-2\rmi\frac{\Delta\omega(H)}{\omega}\right)
\end{equation}
\noindent where $G_s$ is a geometric factor of the resonator.\\
In the spirit of extracting as much information as possible from the raw data, we first report and discuss the measured $Q_u$ against $H$ (upper panel of Figure \ref{fig:Qu}), at selected temperatures.

As it can be seen, $Q_u$ quickly decreases even by applying small magnetic fields. This clearly indicates a large magnetoresistance, whose magnitude can be better appreciated by comparison with similar measurements performed on a typical Y-123 film. Indeed, the Y-123 data (shown in the lower panel of Figure \ref{fig:Qu} at the same reduced temperatures of the preceding set) show a much smaller field induced decrease of $Q_u$ \cite{notaY123}. Following conventional approaches \cite{golo}, it is possible to estimate the vortex viscosity $\eta$, for which we find values fully consistent with the literature \cite{golo}. The comparison brings into light a peculiar feature of the measurements on Tl-2212: since the magnetoresistance is upper bounded by the free flux flow resistivity $\rho_{ff}$, the large magnetoresistance in Tl-2212 determines a peculiarly high $\rho_{ff}$ or, equivalently, a small $\eta$. This result arises directly from the raw data.
\begin{figure}[h]
\centerline{\includegraphics[width=8.5cm]{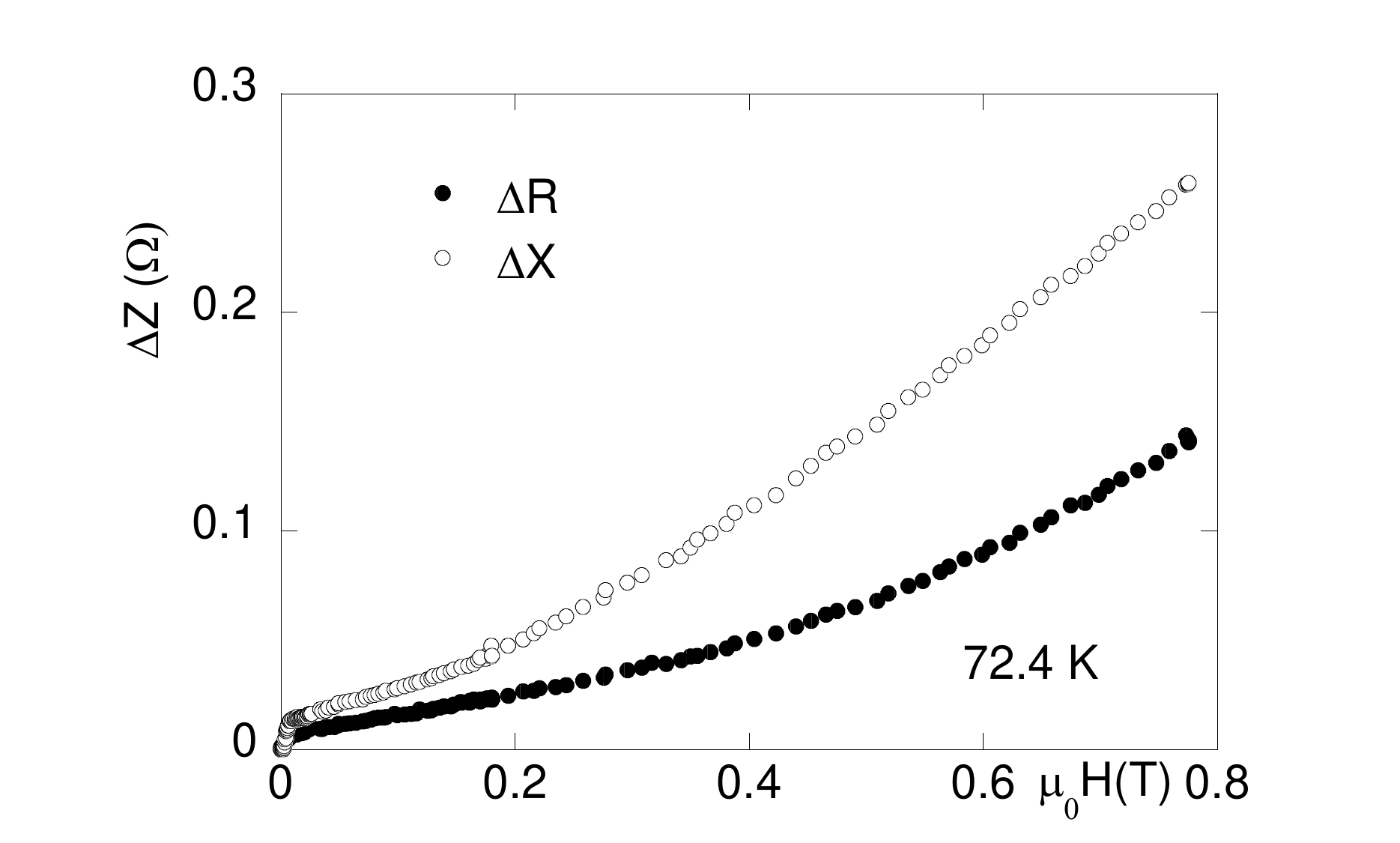}}
\vspace{-3mm}
  \caption{Surface impedance field induced changes $\Delta Z(H)$ for Tl-2212.}
\label{fig:DZ}
\end{figure}

A second feature that is apparent in the comparison presented in Figure \ref{fig:Qu} is the characteristic curvature of the $Q_u(H)$ curves in Tl-2212, which yields additional information on the vortex regime. In fact, when the data are presented as the complex impedance shift (from Eq. \ref{eq:measrho}), we observe a peculiar upward curvature and a strong reactive contribution $\Delta X(H)>\Delta R(H)$. An illustrative set of data for $\Delta Z(H)$ is reported in Figure \ref{fig:DZ} (additionally, a small, quasi-step-like feature can be noted at low fields, presumably related to grain boundaries response; in the remainder of this work, the very low field range will be neglected). Since in thin films one has \cite{thin} $\Delta Z(H)=\Delta\tilde\rho(H)/d$, and (in absence of a significant field-induced pair breaking) $\Delta\tilde\rho(H)\approx\rho_{vm}(H)$, the experimental result of a superlinear field dependence of $\rho_{vm}$ demonstrates that the vortex parameters ought to be field dependent because the resistivity is not proportional to the flux density \cite{london}.\\
%
Within this framework and by resorting to the Brandt model (Equation \ref{eq:rhoc}), the ratio $\Delta X/\Delta R$ yields a lower bound for the normalized pinning frequency, so that $\omega_p/\omega>1$: the observed large reactance involves, therefore, a strong elastic contribution in the vortex dynamics. In addition, this elastic term decreases as $T$ is increased.
In order to extract also all the other fluxon parameters, the creep contribution cannot be a-priori neglected. This fact, together with the field dependence of the vortex parameters, makes the full analysis a quite complex task.
%
However, by deeply exploiting the physical and algebraical properties of the model, $k_p$ and $\eta$ can be evaluated within proper bounds by taking into account the uncertainty on the creep factor. The full treatment is quite lengthy \cite{talliotech}. We mention and use here only the relevant result that, as long as $\Delta X/\Delta R>1$, the pinning constant $k_p(H)$ can be determined from the data for $\Delta\tilde\rho(H)$ with vanishingly small uncertainty, (the resulting $k_p$ comes out to be very close to the one that might be obtained through the simpler GR approach, and almost independent on the values that the creep factor can attain) and upper and lower bounds for $\eta$ never differ more than a factor of 2. The case $\Delta X/\Delta R>1$ applies to our measurements in Tl-2212, not too close to the transition.
\begin{figure}[h]
\centerline{\includegraphics[width=8.5cm]{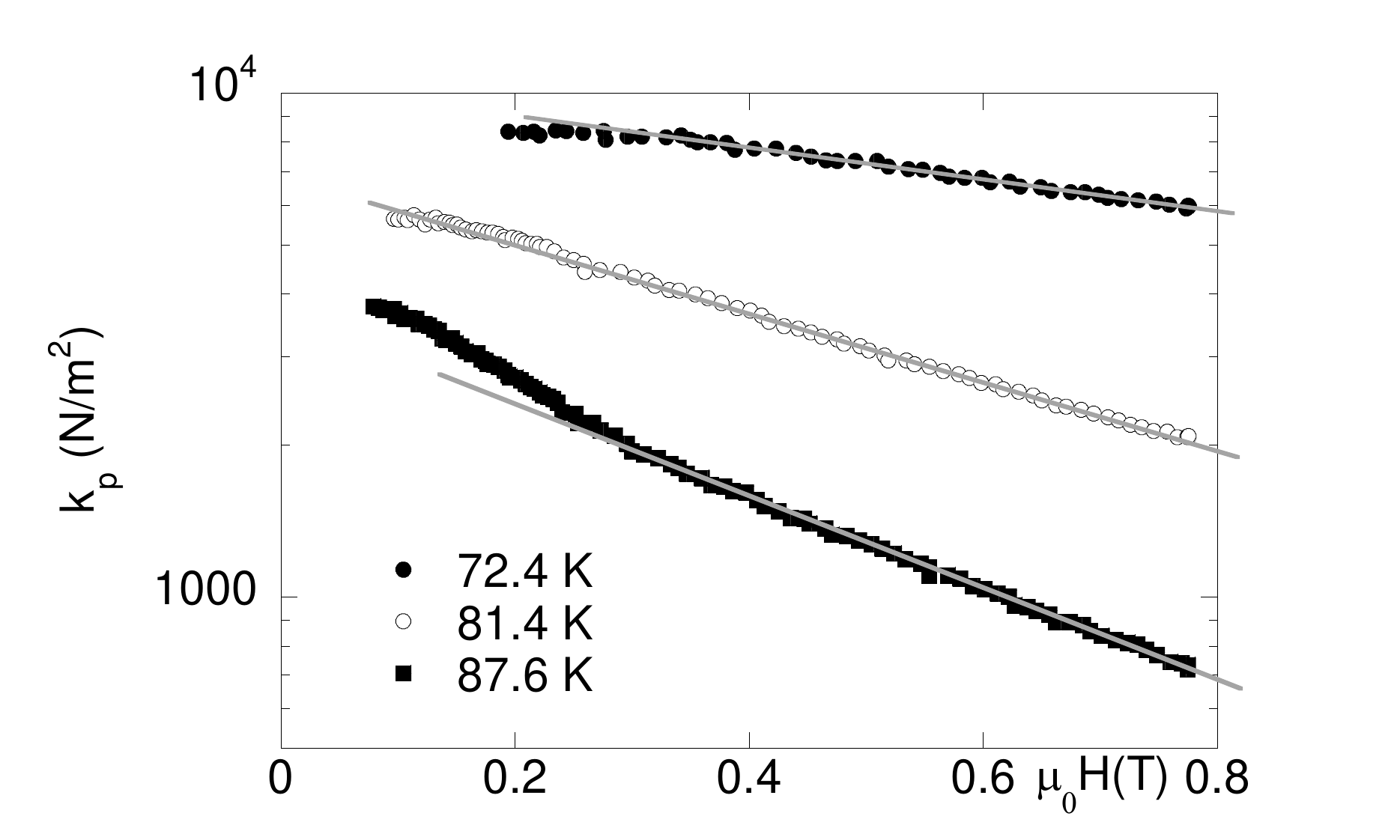}}
\vspace{-3mm}
  \caption{Pinning constant $k_p vs. H$ at selected $T$ for Tl-2212. $k_p$ smoothly decreases by increasing $H$ and $T$.}
\label{fig:kp}
\end{figure}

Results for $k_p$ at selected temperatures are reported in Figure \ref{fig:kp} (the low field region is not shown because of the appearance of the small low-field feature that we ascribe to grain boundaries):  $k_p$ exhibits an exponential decrease with the field, as depicted by the thin lines through the data in Fig. \ref{fig:kp}. We checked that $k_p$ cannot be fitted by any reasonable power law $k_p\sim H^{-\alpha}$. We propose the following interpretation. In a superconductor with strong point pinning and flexible vortices (as expected in Tl-2212 due to the strong anisotropy) the pinning of a single vortex by multiple pins (collective pinning regime) gives rise to a pinning constant dominated by the line tension \cite{blatterone,golo}, which in its turn is essentially dictated by the $c_{44}$ elastic modulus. The latter decreases exponentially with the field when Josephson coupling is the main origin of the vortex tension \cite{daemen}. Thus, our data suggest that the weakening of the elastic response in Tl-2212 is due to a gradual (no abrupt transitions are observed) softening of Josephson-coupled vortices, eventually reaching a full decoupling at higher fields.

Once $k_p$ is determined, one can calculate the allowed range for the vortex viscosity, $\eta_{min}<\eta<\eta_{max}$. The resulting data are compatible with the usually expected field independence \cite{golo}. We stress that, within the model adopted, the values of $\eta$ cannot vary outside the given range. The numerical values for $\eta$, along with $k_p$ evaluated at 0.2 T, are summarized in Table \ref{tab:valori}, and they confirm the initial remark about the observed large magnetoresistance, indicating that the vortex viscosity in Tl-2212 is an order of magnitude smaller than typical values in Y-123 compounds \cite{golo}, but compatible with the low-temperature estimate $\eta\sim$ 5~10$^{-8}$Ns/m$^{2}$ in Bi-2212 \cite{maedaB} (a comparison with Bi-2212 is more difficult, since a very few measurements of the vortex viscosity in the same dynamic range exist).
\begin{table}
\begin{center}
\begin{tabular}{| c | c | c | c |}
\hline
~$T (K)$~ & ~$\eta_{min}$(Ns/m$^{2}$)~ & ~$\eta_{max}$(Ns/m$^{2}$)~ & ~$k_p$(N/m$^{2}$)~\\
 \hline
72.4 & 6.1~10$^{-9}$ & 12~10$^{-9}$ & 8400\\
\hline
81.4 & 3.4~10$^{-9}$ & 6.6~10$^{-9}$ & 5200 \\
\hline
87.6 & 2.4~10$^{-9}$ & 4.3~10$^{-9}$ & 2700 \\
\hline
\end{tabular}
\end{center}
\caption{Numerical values of vortex parameters at selected temperatures. $k_p$ is evaluated at 0.2 T.}
\label{tab:valori}
\end{table}
The physical origin of such small values for $\eta$ is not clear, but almost certainly involves the unconventional nature of the electronic states in cuprates. We propose here some possible speculations. Since the viscosity arises from QP scattering, a reduction of the number of the charge carriers involved or of their scattering rate would result in a reduction of the vortex viscosity. Small density of states within the vortex core could arise as a consequence of the pseudogap opening in close proximity of $T_c$.
Moreover, nodal QP can be largely delocalized so to become less sensitive to the effect of moving vortices \cite{kopnin97}, leading ultimately to a reduced viscosity. This effect can be further enhanced whenever inhomogeneous QP lifetimes on the Fermi surface can be invoked \cite{geshkenbein}. Measurements for different dopings or on samples with different QP lifetime would aid in testing those speculations.


\section{Summary}
\label{conc}
We have presented microwave measurements of the complex resistivity $\tilde\rho$ field induced changes in Tl-2212 thin film. Large magnetoresistance, high reactive contributions as well as peculiar superlinear field dependence of $\tilde\rho$ find a description in terms of the models for vortex dynamics. Macroscopic vortex parameters were extracted. The small vortex viscosity suggests nonconventional  vortex cores. The exponential field decrease of the pinning constant $k_p$ is interpreted as a manifestation of the field-induced weakening of the fluxon stiffness, pointing to a progressive evolution of the vortex system toward decoupled pancakes.
\\
\\
We thank G. Celentano for having kindly provided the Y-123 sample used in the comparison.




\begin{thebibliography}{00}
\bibitem{notaY123} In thin films the slope of the measured $Q_u(H)$ is inversely proportional to the film thickness. Our Y-123 film is thinner than the Tl-2212 film, reinforcing further the statement that the field-induced losses are much larger in Tl-2212.
\bibitem{london} strictly speaking, this statement is true within the London approximation $B\simeq\mu_0H$. Following \cite{brandtcritical} we have checked using the measured $J_c$ that, above $\mu_0H\simeq$ 50 mT, this is indeed the case.
\end{thebibliography}
\end{document}